\input harvmac
\input epsf

\newcount\figno
\figno=0
\def\fig#1#2#3{
\par\begingroup\parindent=0pt\leftskip=1cm\rightskip=1cm\parindent=0pt
\baselineskip=12pt
\global\advance\figno by 1
\midinsert
\epsfxsize=#3
\centerline{\epsfbox{#2}}
\vskip 14pt

{\bf Fig. \the\figno:} #1\par
\endinsert\endgroup\par
}
\def\figlabel#1{\xdef#1{\the\figno}}
\def\encadremath#1{\vbox{\hrule\hbox{\vrule\kern8pt\vbox{\kern8pt
\hbox{$\displaystyle #1$}\kern8pt}
\kern8pt\vrule}\hrule}}

\overfullrule=0pt

\noblackbox
\parskip=1.5mm

\def\Title#1#2{\rightline{#1}\ifx\answ\bigans\nopagenumbers\pageno0
\else\pageno1\vskip.5in\fi \centerline{\titlefont #2}\vskip .3in}

\font\caps=cmcsc10

\noblackbox
\parskip=1.5mm


\def\npb#1#2#3{{\it Nucl. Phys.} {\bf B#1} (#2) #3 }
\def\plb#1#2#3{{ \it Phys. Lett.} {\bf B#1} (#2) #3 }
\def\prd#1#2#3{{ \it Phys. Rev. } {\bf D#1} (#2) #3 }
\def\prl#1#2#3{{ \it Phys. Rev. Lett.} {\bf #1} (#2) #3 }

\def\bb#1{{\tt hep-th/#1}}

\def\npps#1#2#3{{ \it Nucl. Phys. Proc. Suppl. } {\bf #1} (#2) #3 }

\def\jhep#1#2#3{{ \it J. High Energy Phys.} {\bf #1} (#2) #3 }


           \def\CO{{\cal O}} 
   
\def\CL{{\cal L}}


\def\dj{\hbox{d\kern-0.347em \vrule width 0.3em height 1.252ex depth
-1.21ex \kern 0.051em}}

\def\half{{1\over 2}\,}

\def\tp{\tilde p}

\def\Dirac{\,\raise.15ex\hbox{/}\mkern-13.5mu D}
\def\dirac{\,\raise.15ex\hbox{/}\kern-.57em \partial}
\def\shalf{{\ifinner {\textstyle {1 \over 2}}\else {1 \over 2} \fi}}
\def\sshalf{{\ifinner {\scriptstyle {1 \over 2}}\else {1 \over 2} \fi}}
\def\sfourth{{\ifinner {\textstyle {1 \over 4}}\else {1 \over 4} \fi}}
\def\Pf{{\rm Pf}}
\lref\rrev{M.R. Douglas and N.A. Nekrasov, \bb{0106048\semi}
R.J. Szabo, \bb{0109162.}}

\lref\rdipoles{C.-S. Chu
and P.-M. Ho,
\npb{550}{1999}{151} \bb{9812219\semi}
 M.M. Sheikh-Jabbari, \plb{455}{1999}{129} \bb{9901080\semi}
D. Bigatti and L. Susskind, \prd{62}{2000}{066004}  \bb{9908056.}
}

\lref\rezawa{Z.F. Ezawa, {\it ``Quantum Hall Effects"}, World Scientific (2000).}

\lref\rpr{R.D. Pisarski and S. Rao, \prd{32}{1985}{2081.}}

\lref\rjt{S. Deser, R. Jackiw and S. Templeton, {\it Ann. Phys. (N.Y.)} {\bf
140} (1982) 372.}

\lref\ragbar{L. Alvarez-Gaum\'e and J.L.F. Barb\'on, \bb{0109176.}}

\lref\rka{A. Gorsky, I.I. Kogan and C. Korthals-Altes,
\bb{0111013.}}

\lref\rdbranes{O. Bergman, Y. Okawa and J. Brodie, \jhep{11}{2001}{019}
\bb{0107178\semi}
S. Hellerman and  L. Susskind,
\bb{0107200.}}

\lref\rlawv{R.B. Laughlin, \prl{50}{1983}{1395.}}

\lref\ruvir{S. Minwalla, M. Van Raamsdonk and N. Seiberg, \jhep{02}{2000}{020}
\bb{9912072.}}

\lref\rsut{
M. Hayakawa,
\plb{478}{2000}{394} \bb{9912094,}
 \bb{9912167\semi}
A. Matusis, L. Susskind and N. Toumbas, \jhep{12}{2000}{002}
\bb{0002075\semi}
I.Ya. Aref'eva, D.M. Belov and A.S. Koshelev, \npps{102}{2001}{11} 
\bb{0003176.}}

\lref\rchucs{C.-S. Chu, \npb{580}{2000}{352}  \bb{0003007.}}

\lref\rshift{G.-H. Chen and Y.-S. Wu, \npb{593}{2001}{562} \bb{0006114.}}

\lref\redge{A. Pinzul and A. Stern, \jhep{11}{2001}{023} \bb{0107179 \semi}
A.R.  Lugo,
\bb{0111064.} }

\lref\rswex{H. Liu, \npb{614}{2001}{305} \bb{0011125\semi}
Y. Okawa and H. Ooguri, \prd{64}{2001}{046009} \bb{0104036\semi}
S. Mukhi and N.V. Suryanarayana, \jhep{05}{2001}{023} \bb{0104045\semi}
H. Liu and J. Michelson, \plb{518}{2001}{143} \bb{0104139.}}

\lref\rksak{D. Karabali and B. Sakita,
\bb{0106016.}}

\lref\rzhang{D.-H. Lee and S.-C. Zhang, \prl{66}{1991}{1220.}}

\lref\rgirvin{S.M. Girvin, A.H. Mac-Donald, P.M. Platzman, \prl{54}{1981}{581.}}

\lref\rstone{M. Stone, {\it Phys. Rev.} {\bf B42} (1990) 212.}

\lref\rread{N. Read, \prl{62}{1989}{86.}}

\lref\reffac{M. van Raamsdonk, \jhep{11}{2001}{006} \bb{0110093\semi}
 A. Armoni and E.   L\'opez,
\bb{0110113.}}

\lref\rsus{L. Susskind, \bb{0101029.}}

\lref\rpold{A.P. Polychronakos, \jhep{04}{2001}{011}
 \bb{0103013,} \jhep{06}{2001}{070}
\bb{0106011.} }

\lref\rgrandi{N. Grandi and G.A. Silva, \plb{507}{2001}{345}  \bb{0010113.}}

\lref\rhelram{S. Hellerman and  M. Van Raamsdonk,\jhep{10}{2001}{039}
\bb{0103179.}}

\lref\rcsqh{S.-C. Zhang, T.H. Hansson and S. Kivelson, \prl{62}{1988}{82.}}

\lref\rsusb{S. Bahcall and L. Susskind, {\it Int. J. Mod. Phys.} {\bf B5}
(1991) 2735.}

\lref\rkalhal{C. Kallin and B.I. Halperin, {\it Phys. Rev.} {\bf B30}
(1984) 5655.}

\lref\rSW{N. Seiberg and E. Witten, \jhep{09}{1999}{032,} \bb{9908142.}}


\baselineskip=15pt

\line{\hfill CERN-TH/2001-374}
\line{\hfill US-FT-14-01}
\line{\hfill {\tt hep-th/0112185}}

\vskip 0.2cm

\Title{\vbox{\baselineskip 12pt\hbox{}
 }}
{\vbox {\centerline{Noncommutative Field Theory and the Dynamics}
\vskip10pt
\centerline{of Quantum Hall Fluids}
}}

\vskip0.3cm

\centerline{$\quad$ {\caps
J.L.F. Barb\'on\foot{On leave
from Departamento de F\'{\i}sica de Part\'{\i}culas da
 Universidade de Santiago de Compostela, Spain.} }}

\vskip0.1cm

\centerline{{\sl  Theory Division, CERN,
 CH-1211 Geneva 23, Switzerland}}
\centerline{{\tt
barbon@cern.ch}}

\vskip0.3cm

\centerline{$\quad$ {\caps
A. Paredes
 }}
\vskip0.1cm

\centerline{{\sl Departamento de F\'{\i}sica de Part\'{\i}culas.
Universidade de Santiago}}
\centerline{{\sl E-15706 Santiago de Compostela, Spain}}
\centerline{{\tt
angel@fpaxp2.usc.es}}

\vskip0.2cm

\centerline{\bf ABSTRACT}

 \vskip 0.1cm

 \noindent

We study the spectrum of density fluctuations of Fractional  Hall Fluids 
in the context of the noncommutative hidrodynamical model of Susskind.
We show that, within the weak-field expansion, the leading correction
to the noncommutative Chern--Simons Lagrangian (a Maxwell term in the
effective action,) destroys the incompressibility of the Hall fluid
due to strong UV/IR effects at one loop. We speculate on possible
relations of this instability with the transition to the Wigner crystal,
and conclude that calculations within the weak-field expansion must
be carried out with an explicit ultraviolet cutoff at the noncommutativity
scale.

We point out that the noncommutative dipoles exactly match the spatial
structure of the Halperin--Kallin quasiexcitons. Therefore, we propose that
the noncommutative formalism must  describe accurately the spectrum
at very large momenta, provided no weak-field 
 approximations are   made. We further conjecture that the noncommutative
open Wilson lines are `vertex operators' for the  
quasiexcitons.

\vskip 0.1cm

\Date{December 2001}


\vfill





\baselineskip=14pt

\newsec{Introduction}

\noindent

Noncommutative Geometry is relevant to the physics of the Quantum Hall Effect
(QHE) in various guises. The most fundamental is the fact that projection
to the lowest Landau level (LLL) yields non-commuting position operators
for planar electrons in a magnetic field, i.e.
\eqn\ncg
{\left[ X, Y \right] = i\;\theta_B
}
where $X,Y$ denote the projected operators and
\eqn\thetab
{\theta_B = {\hbar c  \over eB}
.}
The area of an elementary quantum of magnetic flux is given by
$2\pi\theta_B$.
  Equation \ncg\ largely determines the properties of free electrons in
the LLL, and
 leads to the natural appearance of the Moyal algebra and non-relativistic
noncommutative field theory (NCFT) with deformation parameter $\theta_B$
 (c.f. for example \refs\rezawa).
 Notice
that this nocommutative geometry of the LLL arises dynamically at the quantum
level, in particular $\theta_B \rightarrow 0$ in the classical limit.

Recently, Susskind has proposed a different application of NCFT (c.f.
\refs\rrev) in the context
of an effective theory of the incompressible quantum Hall fluid. The proposal
is based on the hidrodynamical model of the  Hall liquid \refs{\rsusb,
\rsus}. In the Lagrangian
point of view, one works with a comoving frame $y^i$ that labels the particles
in the fluid. The flow is represented by functions $x^i (y,t)$ that embed
the $y$-plane (the world-sheet parameters) into real space. A crucial
point is that the statistical permutation group of {\it identical} particles
is classically modelled by area-preserving mappings in $y$-space. Thus, we
are naturally led to a gauge theory of the group of area-preserving
 diffeomorphisms SDiff. The gauge-field picture is literal in terms
of the `displacement field'
\eqn\dff{
\theta^{ij} a_j = x^i - y^i,}
where $\theta^{ij} = \theta\,\epsilon^{ij}$ and $\theta$  is a constant with
dimensions of area that fixes the normalization of the fluid gauge
field. The SDiff group of gauge transformations acts on
the gauge field as:
\eqn\act{
\delta a_j = \partial_j \lambda + \{a_j, \lambda\},}
where $\{\;, \}$ represents the Poisson bracket:
\eqn\pb{
\{f,g\} = \theta^{ij}\,\partial_i f\,\partial_j g.}

In more detail, let us  start from an effective particle Lagrangian
\eqn\mila{
\CL_{\rm particle} = \sum_\alpha \half m_* \,({\dot x}^i_\alpha)^2 -
 {eB\over 2c} \sum_\alpha
\epsilon_{ij}\, {\dot x}^i\, x^j - V_C,
}
with $\alpha$ denoting the many-particle label.
In the microscopic theory, the
Coulomb potential $V_C$ features both  direct and exchange terms,
 of which only the
first one has a standard hidrodynamical (classical) interpretation:
\eqn\coulp{
V_C^{\rm direct}  = {e^2 \over 2\,\epsilon}
 \int d^2 x\; d^2 x'\, (\rho(x)-\rho_0) \,{1 \over |x-x'|}
 \,(\rho(x') - \rho_0)
,}
where $\epsilon$ denotes the dielectric constant and we have normalized
the energy to that of a fluid with ground-state uniform density $\rho_0$.

The basic dynamical hypothesis is made that the interplay between the
Coulomb interactions and the fermionic statistics of the electrons results
in an effective potential that, at least on long wavelength scales,
can be approximated by an ultralocal
 harmonic term,
\eqn\ultral{
V_C \longrightarrow
{\mu\over 2} \int d^2 y\,(\rho(x) - \rho_0)^2.}
As part of this dynamical hypothesis, it is also assumed that
 an effective inertial parameter $m_*$ is generated  as a result of the
same underlying dynamics that leads to \ultral,
c.f.  \refs\rstone.
 Thus, we have an effective kinetic energy for the `dressed electrons' of the
fluid, even if the LLL projection has quenched the kinetic energy of
the bare electrons.

Hence, $\mu$ and $m_*$ are phenomenological parameters  encoding the
Coulomb-force dynamics that generates the gap in the fractionally filled
lowest
Landau level. A derivation of these parameters in terms of the true microscopic
parameters, such as the electron mass and charge, is highly
 non-trivial and
should proceed along the lines of \refs\rread. We now show that the
fluid limit of \mila\ and \ultral\ imply a gapped spectrum with the
standard phenomenology of the QHE.

Passing to the fluid picture one has
\eqn\fden{
\rho(x) = \rho_0 \,|\partial y /\partial x| = {\rho_0 \over 1+\theta^{ij}
f_{ij}} ,}
where $f_{ij} = \partial_i a_j - \partial_j a_i + \{a_i, a_j \}$ is the
Poisson field strenght. Thus, we have a highly non-linear action for the
Poisson gauge theory in the temporal gauge $a_0 =0$. We can restore covariant
notation by interpreting the vortex-free flow conditions as a vacuum Gauss law.
The resulting  field $a_\mu$ becomes a
  $U(1)$ gauge field in the linearized approximation.
 Furthermore,  in the limit of large magnetic field
the leading terms
 in the action at long distances  give a Maxwell--Chern--Simons
Lagrangian
\eqn\cs{
S_{\rm eff} \rightarrow -{1\over  4g^2} \int dt\, d^2 y \,
\left({1\over c_s^2} |f_{0i}|^2 - |f_{ij}|^2 \right)   -  {\hbar k \over 4\pi }
 \int dt \,d^2 y\,\epsilon^{\mu\nu\rho}
\,a_\mu \partial_\nu a_\nu,}
where the velocity of  sound is:
\eqn\vsou{
c_s^2 = {\mu \,\rho_0 \over m_*},}
and the coupling parameters are given by
\eqn\cscou{
g^2 ={1\over \mu \,\rho_0^2 \,\theta^2}, \qquad
  k= {2\pi \,e\, B \,\rho_0 \over \hbar\, c}\, \theta^2.}
  The standard phenomenology of the FQHE then follows
for the particular choice
\eqn\pcho{
\theta = {1\over 2\pi\rho_0} = k\,\theta_B,}
with $1/k = \nu$, the filling fraction.  It is remarkable that this
determination of $\theta$ is formally independent of $\hbar$.
Indeed, the Poisson structure \pb\ makes sense even as a purely classical
description of the fluid.

The spectrum of density fluctuations at long wavelengths is
gapped at the effective cyclotron frequency
\eqn
\cif{
\omega_0 = {g^2 \, c_s^2 \,\hbar\, k \over 2\pi} = {e B \over m_* c}.}
Physically, this gap is mostly induced by Coulomb interactions, i.e.
$\hbar \omega_0 = \CO(e^2 / \epsilon \sqrt{\theta})$. 
 The relation \cif\
 determines phenomenologically
the value of the effective inertial parameter $m_*$.
If the microscopic dynamics is such that $m_* =\infty$, or $k=0$,
 the massless `photon' is
nothing but the phonon of the
acustic
excitations in a superfluid phase. In the Hall phases there is no
such acustic branch of phonons due to the magnetically induced
Chern--Simons mass.  The dispersion
relation following from \cs,
\eqn\dispre{
\omega({\bf p}) = \sqrt{\omega_0^2 + c_s^2 |{\bf p}|^2},}
is phonon-like,
 $\omega({\bf p}) \approx c_s |{\bf p}|$, at very large momenta, but this turns
out to be an unphysical feature of \cs, that should be analyzed in
a low-energy expansion around zero momentum, corrected by a tower of
non-renormalizable operators that are suppressed by powers of $\theta$. In
particular, such operators were discarded in obtaining \cs\ as part
of the weak-field expansion. Thus, \cs\ approximates the physics only
in the regime
\eqn\reg{
|{\bf p}| \ll {\rm min}\;\;(1/\sqrt{\theta},\; \omega_0 /c_s).}

The proposal of ref. \refs\rsus\ consists in approximating the statistics
group by a   version
of $U(\infty)$ rather than SDiff,
 i.e. we replace the Poisson brackets by Moyal brackets:
\eqn\sust{
\{f,g\} \rightarrow -i\,[f,g] = -i\,(f\star g - g\star f),}
with the standard definition of the Moyal product:
\eqn\moyp{
f(x)\star g(x) = \lim_{\eta,\xi \to 0} \exp\left({i \over 2} \theta^{\alpha\beta}
{\partial \over \partial \eta^\alpha} {\partial \over \partial \xi^\beta}
\right) \;f(x+\eta) \;g(x+\xi).}
The Moyal and Poisson brackets
differ to second order in the derivative  expansion in powers of
 $\theta^{ij} \partial_i
\partial_j$.
The standard $U(1)$ gauge symmetry is promoted to a noncommutative
$U(1)$ gauge symmetry that completes the long-distance Chern--Simons effective
Lagrangian with nonlinear terms:
\eqn\ncs{
S_{\rm NCCS} = -{\hbar k \over 4\pi} \,\int dt\,d^2 y\,\epsilon^{\mu\nu\rho}
\left(A_\mu \,\partial_\nu A_\rho + {2i\over 3}\,A_\mu \star A_\nu \star A_\rho
\right)
.}
The resulting model succesfully accounts for the connection between the
statistics of the electrons and the filling fraction \refs\rsus.
It is worth stressing at this point the conceptual difference between
the LLL noncommutativity parameter $
\theta_B$, essentially a single-particle effect, and the noncommutativity
of the Lagrangian coordinates $[y^i, y^j] = i\epsilon^{ij} \,\theta$, tied
to the implementation of the quantum statistics of the {\it many body}
system. Numerically, they are related via the filling fraction $\theta_B =
\nu \,\theta$, so that it would be desirable to have a more physical
explanation of the interplay between these two notions of noncommutativity.

A very important aspect of \ncs\ is the fact that, when
 written in terms of the
`electron' coordinate field
\eqn\ef{
X^i = y^i + \theta^{ij}\,A_j,}
it is just a Chern--Simons matrix model where the $X^i$ are
infinite-dimensional matrices whose eigenvalues characterize the individual
electron positions. Therefore, the noncommutative model of the
`fluid' is in principle
capable of capturing the `granularity' of the electrons.
More specific studies of this model
 have proceeded by  truncation to a localized
 finite
electron system (a droplet).
 In this case, the noncommutative Chern--Simons model
collapses to  a  quantum mechanical $U(N)$ Chern--Simons matrix model
for a system of $N$ electrons, together with extra `edge' degrees of freedom
in the fundamental representation \refs\rpold. Formally, this corresponds to
substituting the statistical permutation group  $S_{N}$ by the unitary group
$U(N)$. Notice that the number of gauge-invariant degrees of freedom is the
same, at least in the large-$N$ limit,
 since coordinates of electrons are promoted to hermitian matrices; in the
unitary gauge one has the set of eigenvalues, and the residual
 Weyl group yields the
standard statistics group. The effectiveness of the
finite matrix models
must be judged against the Laughlin wave functions \refs\rlawv, since one is
working directly with $N$ electron degrees of freedom (see \refs{\rhelram,
\rpold,\rksak}.) Thus, it is a more
 microscopic
approach in nature. In this note we will mostly concentrate on the
fluid picture and seek a description of the physics in terms of
continuum field theory.

Both the noncommutative fluid model \ncs\ and its matrix approximations
can be obtained as a Seiberg--Witten limit \refs\rSW\ of certain bound
states  of D-branes
\refs\rdbranes.

\newsec{Beyond the Topological Dynamics}

\noindent

The non-linear completion of the standard Chern--Simons Lagrangian
that is given in \ncs\ is of little  relevance for the bulk physics
of the fluid, for it can be
shown \refs\rgrandi\  that the Seiberg--Witten map to ordinary gauge fields
cancels the non-linear terms and leaves out the standard Chern--Simons
Lagrangian
\cs. This manipulation requires discarding total derivatives, i.e. working
on ${\bf R}^2$ or a manifold without boundary.

A simple proof of this important property can be given by using the
exact solution of the Seiberg--Witten transform found in \refs\rswex. Since
these exact solutions are simple for the field-strength operators, we extend
the gauge bundle on $M_\theta \times {\bf R}$, with $M_\theta$ the
noncommutative plane or the noncommutative torus,  to a bundle on $M' =M_\theta
\times {\bf R} \times [0,1]$ by $A(s) = s \,A$, where $s\in [0,1]$, a {\it
commutative} interval. Then
the Chern--Simons  action on $M_\theta \times {\bf R}$ can be obtained
as a boundary term from the four-dimensional topological action on the
extended space,
\eqn\exac{
S_{\rm NCCS} = -{\hbar k \over 4\pi} \int_{M'} F \wedge F.}
Now, the Seiberg--Witten map of the  ordinary $U(1)$
Lagrangian density in momentum space
is given
 by:
\eqn\swrel{
(f \wedge f) (p) = \int_{M'}
L_* \left[\sqrt{{\rm det}(1-\vartheta F)} \;F {1 \over
1-\vartheta F} \wedge F {1 \over
1-\vartheta F} \;e^{ip(y+ \vartheta A)} \right]
.}
where $L_*$ is the instruction of path-ordering with respect to the
Moyal product and matrix notation in Lorentz indices is understood. Since
we are looking at the integrated Lagrangian density we just need the
zero-momentum coupling that eliminates the path-ordering prescription.
We take $\vartheta^{23} =\theta$ as the only nonvanishing entry of the
noncommutativity matrix.
 By explicit matrix algebra one finds
$
(\vartheta F)^2 = -c\,(\vartheta F)
,$
where $c= \theta F_{23}$. So that
$$
F\,{1 \over 1-\vartheta F} = {1\over 1+c} \,(F+\delta F)
,$$
where $\delta F$ has only non-vanishing entries in the $01$ plane:
$$
(\delta F)_{01} = -(\delta F)_{10} = \theta \,\Pf\,(F)
 ={\theta\over 4} \epsilon^{\mu\nu\rho\sigma}\, F_{\mu\nu}\, F_{\rho\sigma}
.$$
Finally, using
$
{\rm det} (1-\vartheta F) = (1+c)^2
$
and the property
$
\Pf\,(F+\delta F) =
  (1+c) \Pf\,(F)
$, we obtain
$$
\int_{M'} f \wedge f =\int_{M'} \Pf \,(F)= \int_{M'} F\wedge F
$$
for this particular noncommutativity matrix. Picking out the boundary
terms this extends to the Chern--Simons actions in 2+1 dimensions.

The on-shell triviality of the noncommutative Chern--Simons action
  agrees with the
results of \refs\ragbar\ (see also \refs\rka,) where it was found
 that on a spatial torus the model is T-dual to an
ordinary non-abelian (twisted)
 $U(N)$ model whose physical Hilbert space only samples
the diagonal $U(1)$ subgroup. Therefore, the on-shell
 bulk properties of \ncs\
are equivalent to those of the ordinary, topological Chern--Simons
Lagrangian. In hindsight, the difficulties found in deriving out of \ncs\
the conformal field theory of edge states \refs\redge, can be
trivially solved if we decide to quantize the theory resulting after
the Seiberg--Witten field redefinition. In principle, this is a valid
option from the physical point of view, but it largely trivializes the
`kinematical' successes  of \ncs.

Therefore, in order to judge the real impact of  $U(\infty)$  as
a `statistical' symmetry group  of the Hall fluid  it would be desirable
to  go beyond
the topological term. A natural option is to study the gapped density
excitations at low momentum
 by keeping the propagating terms in the fluid effective
action. The noncommutative gauge theory of the fluid follows from
the general expressions above by substituting the Poisson gauge field
$a_\mu$ by the noncommutative gauge field $A_\mu$ and the corresponding
field strength $f_{\mu\nu}$ by
 $F_{\mu\nu} = \partial_\mu A_\nu - \partial_\nu A_\mu -i
A_\mu \star A_\nu + i A_\nu \star A_\mu$. Expanding  in powers of the
 field strength we obtain the noncommutative version of
\cs:
\eqn\mcs{
S_{\rm eff} = -{1\over 4g^2} \int |F_{\mu\nu}|^2 - {k \over 4\pi} \int
\epsilon^{\mu\nu\rho} \left(A_\mu \,\partial_\nu A_\rho + {2i\over 3}
A_\mu \star A_\nu \star A_\rho\right).}
where we have chosen units with
$c_s =\hbar =1$.

To the extent that \mcs\ describes a smooth deformation of \cs\
(such as, for example, via the classical Seiberg--Witten map,) we
expect a gapped spectrum of density fluctuations with $\omega ({\bf p})
\rightarrow \omega_0$ as ${\bf p}\rightarrow 0$. On the other hand, we
would like to find new $\theta$-dependent features of the excitation
spectrum at intermediate momentum scales.

At the classical level, these expectations are borne out since
interactions are analytic in $\theta$. The spectrum is gapped at zero
momentum at the scale of the Chern--Simons mass $\omega_0 =M = g^2 k /2\pi$.
 However, it is well-known that
gauge theories present strong UV/IR mixing at one-loop order,
 rendering the theory
non-analytic around $\theta =0$, at least in perturbation theory
\refs{\ruvir, \rsut}.  In the following, we  study the effect on the
collective excitations  of
the  one-loop
UV/IR mixing. After this standard analysis we  offer a physical interpretation
of the results.

\subsec{Collective Excitations at One Loop}

\noindent

The dispersion of neutral collective excitations follows from the
poles of the polarization tensor $\Pi^{\mu\nu}$ in the
quadratic effective action
\eqn\qeff{
S^{(2)}_{\rm eff} = \half \int A_\mu \,\Pi^{\mu\nu}\,A_\nu.}
The constraints of linearized gauge symmetry and rotational symmetry
on the plane constrain  the tensor structure to be of the form:
\eqn\tens{
\Pi_{\mu\nu} = \Pi_e \;(p_\mu \,p_\nu - p^2\,\eta_{\mu\nu}) + i \,M \;\Pi_o\;
\epsilon_{\mu\nu\rho}\, p^\rho + \Pi_{nc} \;\tp_\mu \,\tp_\nu,}
where $\tp^\mu = \theta^{\mu\nu} p_\nu$. In addition to the usual even and odd
parity selfenergies $\Pi_e, \Pi_o$, we have a new tensor structure $\Pi_{nc}$
which is only contributed by non-planar diagrams. Notice that a term
proportional to $\tp^\mu \tp^\nu$ is perfectly compatible with linearized
gauge invariance, since $p^\mu \tp_\mu =0$. All scalar self-energies
are  functions of  the independent variables $p^2$ and $\tp^2$.

The dispersion relation can be found by inverting the polarization tensor.
The physical pole is located at
\eqn\dispr{
p^2\, \Pi_e^2 - M^2 \,\Pi_o^2 - \tp^2 \,\Pi_e \,\Pi_{nc} =0,}
which yields the dispersion relation (we work in the $(+--)$ signature): 
\eqn\diss{
\omega({\bf p}) = \sqrt{{\bf p}^2 + M^2 \,{\Pi_o^2 \over \Pi_e^2} + \tp^2\;
{\Pi_{nc} \over \Pi_e}}.}

In general, the planar part of the perturbation theory is isomorphic to
that of a $SU(N)$ Yang--Mills--Chern--Simons model in the formal limit
$N\rightarrow 1$. Therefore, according to the analysis of \refs\rjt,
\refs\rpr\ the noncommutative theory will be ultraviolet-finite at one loop.  
 Standard dimensional analysis shows that  the
 effective expansion parameter of planar perturbation theory
is  given by $g^2 / |p|$ at high momenta $|p| \gg M$, whereas the finite
tree-level mass of the photon yields
$g^2 / M \sim 1/k$ as the effective expansion parameter
 at low momenta $|p| \ll M$. The careful analysis of \refs\rpr\ obtains
the scaling expected from dimensional arguments  after subtle cancellations
of infrared effects in which the choice of Landau gauge is instrumental.

The planar parts of the self-energy tensors have a low energy expansion of the
form, c.f. \refs\rpr:
\eqn\lex{
\Pi_{\rm planar} = 1+ \CO(g^2 /M) + \CO(g^2 |p| /M^2) = 1+ \CO(1/k) + \CO(|p|
 /k
M).}
The corrections of  $\CO(1/k)$ are responsible for the effective shift
of the Chern--Simons level \refs\rshift.
The contribution of low loop momenta $|q| < M$ to the
non-planar part of the self-energies is similar,
 since Moyal phases are negligible in
this range.
   On the other hand, the extreme ultraviolet contribution to the non-planar
diagrams exhibits the well-known UV/IR mixing.

The UV/IR mixing is based on the fact that Moyal phases tied to loop
momenta introduce an effective ultraviolet cutoff in non-planar diagrams
given by
$\Lambda_{\rm eff} = 1/|\tp| $. Since the one-loop contribution to
the scalar self-energies is proportional to $g^2 \sim M/k$, the maximal
possible degree of divergence of $\Pi_e, \Pi_o$ is
$\CO(\Lambda_{\rm eff}^{-1})$.
Therefore, the UV  non-planar contributions to $\Pi_e, \Pi_o$ at low momenta
are of $\CO(M|\tp|/k)$.

The Lorentz-violating tensor structure $\Pi_{nc}$ is
different. Since it has mass dimension four, the non-planar one-loop diagram
can be of $\CO(\Lambda_{\rm eff}^3)$. Thus, we expect
 $ \Pi_{nc} \sim 1/ |\tp|^3$.
By explicit inspection, one finds that the most singular contribution
 to the diagrams is the integral
\eqn\maxx{
\int {d^3 q \over (2\pi)^3} \,{2\,q^\mu\, q^\nu
 - \eta^{\mu\nu}\,q^2 \over q^4} \;
e^{i\tp\,q} \sim {\tp^\mu\,\tp^\nu \over |\tp|^3} + {\rm less} \;\;\;{\rm
singular}.}
One finds (see Appendix A for a more detailed  discussion):
\eqn\expnc{
\Pi_{nc} = {M \over 2 k} \left({1\over |\tp\,|^3} +\CO(M^2/|\tp\,|) \right).}
This term is the first in a power expansion of the full gauge-invariant
effective action that must be written in terms of open Wilson lines,
as in \refs\reffac.
A very significative fact is that the leading term comes with a {\it positive}
sign. Thus, plugging this back into the general expression for the dispersion
relation, one finds that, at low momenta:
\eqn\insd{
\omega({\bf p})^2 = {\bf p}^2 + M^2 -{M\over 2 k} \,{1\over \theta \,
|{\bf p}|} + \dots}
The dots stand for neglected contributions of  $\CO(1/k)$, $\CO(|{\bf p}|/k M)$
and $\CO(M^3\, \theta \,|{\bf p}|/k)$. Defining the dimensionless quantity
\eqn\ddelta{
\Delta = {1\over M^2 \,\theta},}
we can determine the range of applicability of \insd\ to be $|{\bf p}|/M
\ll \Delta$.

Therefore, we obtain the expected result that noncommutative UV/IR effects
turn the gauge theory unstable in the infrared, despite the presence of
a gauge-invariant Chern--Simons mass. Imaginary frequencies occur
for sufficiently small momenta.
 If this happens for $|{\bf p}| \ll M$, the critical momenta
for the onset of tachyons is
\eqn\cit{
{|{\bf p}|_c \over M} \approx {1\over k} \,\Delta.}
The physical interpretation is that perturbation theory breaks down at these
values of the momenta. In principle, the infrared dynamics that resolves
these singularities could be non-perturbative.

\newsec{Physical Interpretation}

\noindent

The one-loop
 instability of the noncommutative Maxwell--Chern--Simons theory is
at odds with the naive expectations based on a local (in powers of momenta)
quantization of the model. In particular, one does not obtain the
zero-momentum gap of density fluctuations that is the landmark of the Hall fluids.
There are basically two possibilities regarding the physical interpretation
of the tachyonic singularity: either it signals a physical instability
of the fluid, or it is an artifact of the approximations used and
has no relation to the physics of the real Hall fluid.

In principle, we should expect some kind of instability in the noncommutative
perturbation
theory, since the effective expansion parameter is $1/k = \nu$, the filling
fraction. For small values of the filling fraction the incompressible
fluid is unstable towards condensation into a solid phase, the so-called
Wigner crystal. If the electron density is too low, each elementary cyclotron
orbit is well separated from the others and the lowest energy state is
an hexagonal two-dimensional lattice held by the Coulomb repulsion. As any
other crystal, it has gapless phonon excitations.

Therefore, it is tempting to identify the tachyonic instability at
\cit\ as a result of the expected physics at small filling fraction,
 $\nu \ll 1$, i.e. we are
forcing the fluid description in a region where the Wigner crystal sets in.
If this interpretation is correct, the behaviour of the dispersion relation
in the intermediate regime of momenta
\eqn\indre{
\nu\,\Delta \ll {|{\bf p}| \over M} \ll \Delta}
is interesting  and could capture some qualitative features of the
real excitation spectrum of the Hall fluid.

The dispersion of the lowest neutral excitations in a real Hall fluid has
the qualitative form depicted in Fig. 1. The lowest energy mode occurs
at a finite value of the momentum $|{\bf p}|_{\rm min} \sim 1/\sqrt{\theta}$,
the so-called magnetoroton, in analogy with the analogous modes in
 superfluids \refs\rgirvin.
 As the filling fraction is reduced, the magnetoroton
eventually becomes unstable and a Wigner crystal is formed with
lattice size of $\CO(\sqrt{\theta})$.  The post-magnetoroton regime
at $|{\bf p}| > 1/\sqrt{\theta}$ has $\omega({\bf p}) \sim 1/|{\bf p}|$
and reflects peculiar physics that we will discuss in the next section.

\fig{\sl Qualitative form of the dispersion relation of a Hall fluid.
The excitation of lowest frequency is the magnetoroton that  
 occurs at the scale of the inter-particle separation
 $|{\bf p}_W| \sim 1/\sqrt{\theta}$ and presages the Wigner crystal.
The post-roton regime at larger momenta  describes  the
so-called `quasiexcitons' and scales as
$1/|{\bf p}|$.
}{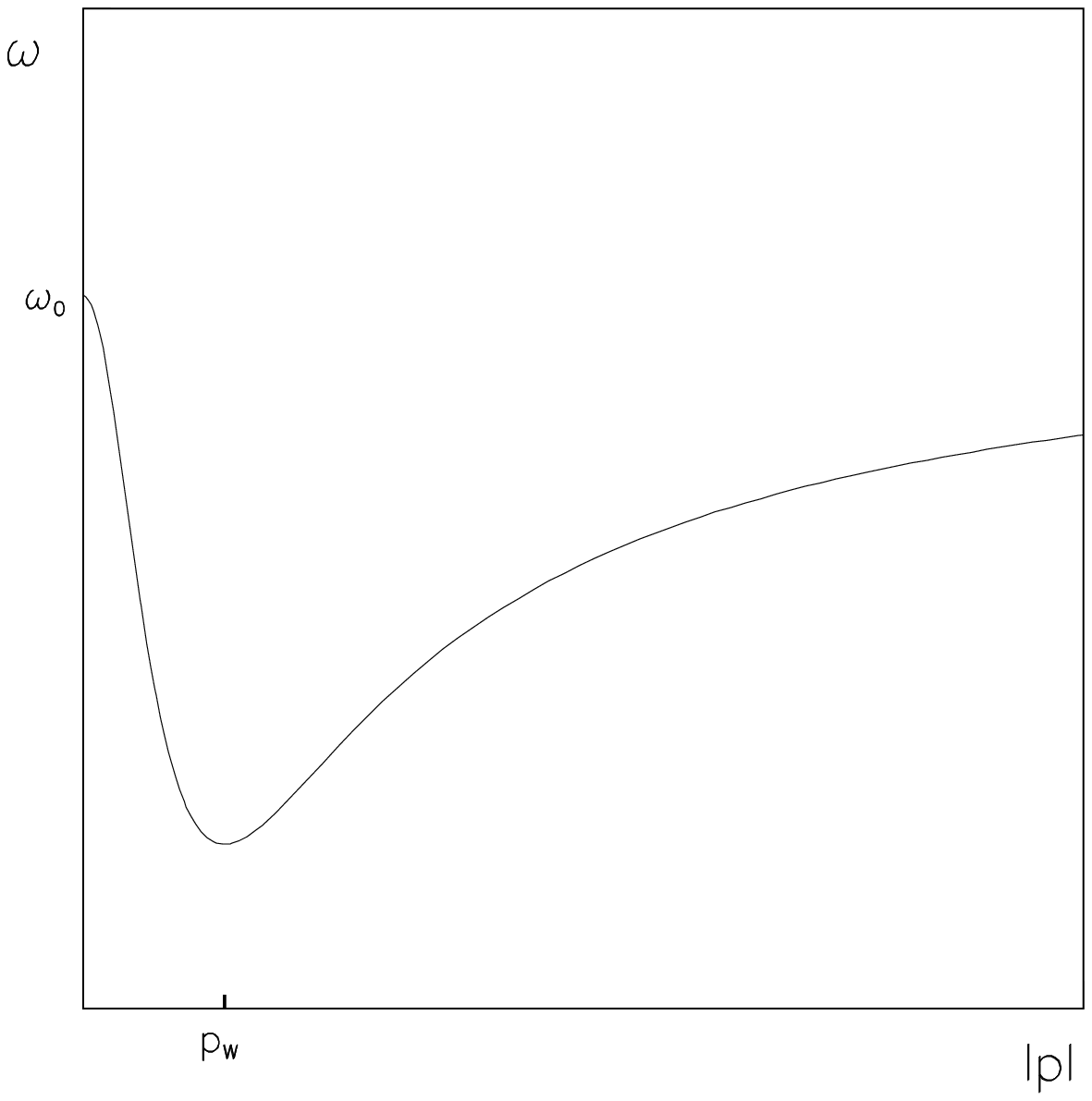}{5truein}

We can formally regularize the infrared singularity
 by artificially turning off the
UV/IR effects. This can be achieved by introducing an explicit ultraviolet
cutoff $\Lambda$ in the loop integrals. The effective cutoff of the
non-planar diagrams then becomes
\eqn\efcut{
\Lambda_{\rm eff}^2 = {1\over |\tp|^2 + 1/\Lambda^2}.}
The effect of this modification in the expression \insd\ is to turn
the dispersion curve back up at momenta of order $|{\bf p}| \sim 1/\Lambda
\theta$. Thus the effect of the ultraviolet cutoff is to mimic the
magnetoroton minimum!

In practice, one must ensure that the cutoff procedure  respects
gauge invariance and does not introduce
longitudinal terms in the polarization tensor. Such terms would give
$\Lambda$-dependent additive renormalizations of the frequency at
zero momentum. On general grounds, we know that a cutoff procedure
that respects gauge invariance must  restore
Lorentz-invariance asymptotically
 at momenta of order $|{\bf p}| \ll 1/\Lambda \theta$.
Therefore, the `noncommutative' part of the polarization tensor must
vanish at zero momentum and the long-distance gap is governed by
the Chern--Simons mass $M$. We include in Appendix B a proof of
the stability of the zero-momentum gap by the addition of a 
gauge-invariant Pauli--Villars
regulator.

If the ultraviolet cutoff is chosen at a value of order $\Lambda \sim
M\,k$, the turn-over coincides with the onset of non-perturbative
effects, i.e. $|{\bf p}| \sim 1/\Lambda \theta \sim \nu M \Delta$.
Then, the dispersion curve in the immediate   post-roton regime
$\nu \Delta < |{\bf p}|/M < \Delta$  is approximately given by
\eqn\apppp{
\omega({\bf p}) \approx M\sqrt{1-{\nu \over 2M\theta |{\bf p}|}} \approx
M-{\nu \over 4 \theta |{\bf p}|},}
the correct qualitative behaviour. Therefore, it is tempting to regard
an ultraviolet cutoff at $\Lambda \sim M k$ as a simulation of
the magnetoroton minimum and the one-loop induced term in  \apppp\
as  a `holographic' calculation of the post-roton dispersion
curve.

\fig{\sl
 Dispersion relations of the noncommutative Maxwell--Chern--Simons model,
 obtained with different values
of the ultraviolet cutoff $\Lambda$.
 The solid line represents the curve without
cutoff and its instability at $|{\bf p}|_c$ {\rm \cit}. The dotted line
corresponds to a cutoff of the order the typical noncommutative scale
$\Lambda \sim 1 / \sqrt{\theta}$. In this limit, ordinary Maxwell--Chern--Simons
theory must be recovered. The dashed line represents an intermediate
cutoff that simulates the magnetoroton.
 Notice that  all the curves become
linear
 for $|{\bf p}|> M$. In the application to the Hall fluids, this regime
is unphysical (an artifact of the weak-field expansion). 
}{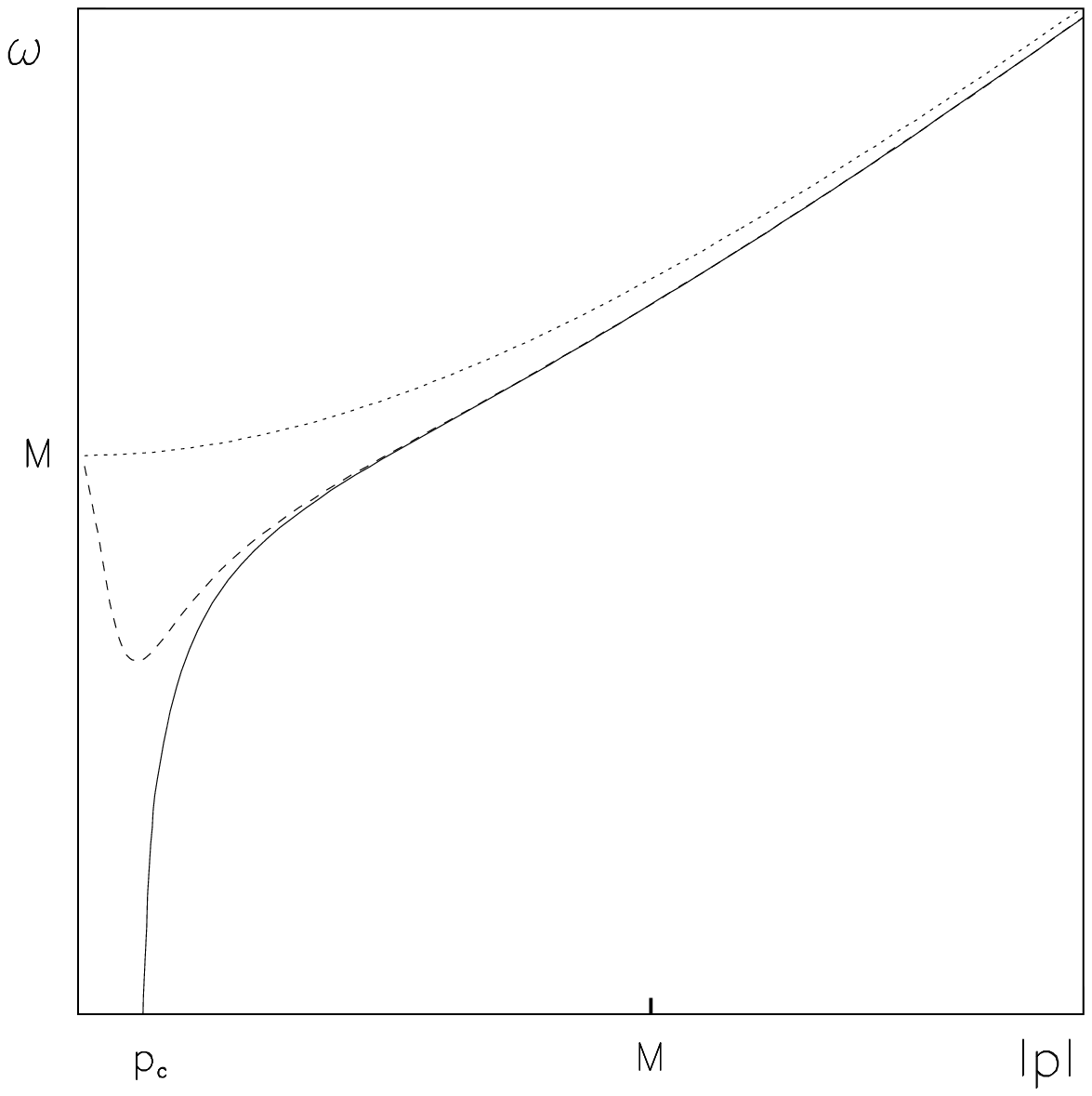}{5truein}

\vskip1cm

Despite its qualitative appeal, the previous picture is probably unphysical
for a variety of reasons.
First of all, the scales do not come out exactly right. If the magnetoroton
minimum is to be associated to perturbative UV/IR effects, one should
have $\sqrt{\theta} M > 1$, since the dispersion curve shows an unphysical  
linear form  for $|{\bf p}| > M$. This constraint implies that $\Delta <1$. But
then we have $\nu \Delta  \ll \Delta < \sqrt{\Delta} = 1/M\sqrt{\theta}$.
Therefore the $\Lambda$-induced minimum at $|{\bf p}| \sim \nu \Delta M$
  always occurs at a much lower
momentum than the expected magnetoroton scale 
$|{\bf p}_W| \sim 1/\sqrt{\theta}$. It is also clear that
the formula \apppp\ always applies at lower momenta than the expected
magnetoroton momentum.

 A more serious problem has to do with the physical specification of
the ultraviolet cutoff. It is plain that the behaviour \apppp\
is induced by the integration over virtual modes of momentum in the range
$M < |{\bf p}| < Mk$. However, this is precisely the regime where the
dispersion relation is approximately linear and thus unphysical! A more
reasonable choice for the scale of the ultraviolet cutoff, such as
$\Lambda \sim M$, eliminates the appealing qualitative behaviour \apppp\
from the dispersion curve.

In fact, since we are neglecting higher powers of $F_{ij}$ in a weak-field
expansion, the natural Wilsonian cutoff for calculations with the
action \mcs\ would be $\Lambda \sim 1/\sqrt{\theta}$, since the weak-field
expansion is organized in powers of  $\theta^{ij} \partial_i \, \partial_j$.
However, in this case the strong UV/IR effects completely dissapear, since
these come from virtual loop momenta in the range $1/\sqrt{\theta} <
|{\bf p}| < \Lambda$. If we take this more physical point
of view, the  quantum corrections at low momentum calculated with
\mcs, equipped with a cutoff at $\Lambda \sim 1/\sqrt{\theta}$,  become
qualitatively indistinguishable   from those obtained by calculation
in the commutative Maxwell--Chern--Simons model. We have summarized this
situation in Fig. 2. 

\newsec{Noncommutative Dipoles and the Halperin--Kallin Quasiexciton}

\noindent

If the noncommutative description of density fluctuations is to capture
some {\it qualitative} features of the real system, it would be good
to identify the basic degrees of freedom via some simple physical
argument. We now come to the basic observation of this paper by providing
such an intuitive    link.

Physically, the main feature of   noncommutative theories  that explains
the peculiarities of their non-local dynamics, including the UV/IR effects,
is the dipolar nature of the elementary excitations. As far as their
interactions are concerned, noncommutative
quanta of momentum $p^\mu$ can be thought of as rigid dipoles of
orientation vector $L^\mu = \tp^\mu = \theta^{\mu\nu} \,p_\nu$ \refs\rdipoles.
This picture is very intuitive in the stringy regularization of
NCFT, in which one starts with open strings. The Seiberg--Witten low
energy limit decouples the normal vibrational modes of the
open strings and the large magnetic field prevents them from shrinking to
a point-like object.

In open string theory, the endpoints are formally oppositely
charged sources for the
massless photon excitation of the open string.
In the fluid description of the QHE, the charged sources for
the fluid gauge field $A_\mu$ are Laughlin's quasiholes and quasielectrons.
Thus, the noncommutative photons could be thought of (at least in certain
regime) as a description of quasiparticle bound states. This intuitive
correspondence is incorporated in the D-brane models of the QHE \refs\rdbranes,
where one explicitly assigns Laughlin quasiparticles to end-points of
open strings on certain D-branes.

An intuitive picture of density fluctuations of the Hall fluid was
developed by Halperin and Kallin \refs\rkalhal\ precisely in terms of
quasiparticle bound states (see also \refs\rzhang.)
 The basic idea is that, since the elementary
gapped excitations about the Laughlin ground state are the quasiparticles,
a neutral density fluctuation can be thought of as a quantization of
quasihole-quasielectron pairs. If these quasiparticle pairs are
well-separated compared to their size, a simple argument gives the
linear extension of the dipole:
in a stationary configuration the Lorentz force must be balanced against
the Coulomb attraction of the pair,
\eqn\balance
{(e \nu) \,{v\over c}\,B = {\partial V_C \over \partial \ell},}
where $\ell$ is the relative separation, $v$ is the velocity in the
orthogonal direction,
and $e\nu$ is the electric charge of the quasiparticles
in the Laughlin fluids with $1/\nu =$ odd. The velocity is related to
the momentum vector by
\eqn\rel{
{\bf v} = {\partial E({\bf p}) \over \hbar\,\partial {\bf p}}.}
Now, in the microscopic system, the LLL projection means that all the
energy of the fluid is well approximated by the Coulomb energy,
$E\approx V_C$. Therefore, we have
\eqn\match{
\ell \approx {\hbar c \over \nu e B} |{\bf p}| = {\theta_B |{\bf p}| \over \nu}
= \theta\,|{\bf p}|.}
We have found that
 the dipolar structure of the Halperin--Kallin  quasiexciton
exactly matches the noncommutative dipole! We think that this is not a
coincidence and actually provides a strong indication that the basic
insight of \refs\rsus\ might be correct.

Based on the quasiexciton picture, one can argue that the large-$|{\bf p}|$
behaviour of the dispersion relation is saturated by the Coulomb potential
between the quasiparticles separated by a distance $\ell \approx \theta |{\bf
p}|$:
\eqn\largep{
\omega({\bf p}) \rightarrow \omega_C({\bf p})= E_{\rm pair} -
{e^2 \,\nu^2 \over \epsilon \,\theta |{\bf p}|}
,}
where $E_{\rm pair}$ is the energy to create a separate pair of
quasiparticles. Notice that the surprising match of the quasiexciton
dipole with the noncommutative dipole does not depend on the particular
value of the interaction potential, in this case the Coulomb potential.
In particular, the Halperin--Kallin argument provides the analog of
the Seiberg--Witten limit for this particular system. The Hamiltonian
 of the free dipoles is therefore given by
\eqn\fhal{
H_{\rm free} \approx  
\int d^2 {\bf p} \;\omega_C ({\bf p}) \;W_{\bf p}^\dagger\;
W_{\bf p},}
 where $W_{\bf p}^\dagger$ creates
a dipole of momentum ${\bf p}$. Clearly, the applicability of noncommutative
perturbation theory to the Hall system depends on whether the interactions
between the quasiexcitons are well-modelled by local
splitting-joining of dipole endpoints, i.e. by quasiparticle pair
creation-annihilation. In this context, the issue of
UV/IR effects must be reexamined with the new propagator based on \fhal.

It is tempting to associate the operators $W_{\bf p}$ to modes of gauge-invariant
open Wilson lines. If this conjecture is true, there must be a very
non-trivial representation of soliton-antisoliton bound states in terms
of open Wilson line `vertex operators'.

For momenta $|{\bf p}| \sim 1/\sqrt{\theta}$, corresponding
to the average inter-electron distance, the picture of well-separated
quasiparticles breaks down, and the screening of charge causes the
  dispersion curve to turn up. The minimum is the so-called magnetoroton
excitation that sits around $|{\bf p}_{\rm min}| \sim 1/\sqrt{\theta}$
and presages the instability towards the Wigner crystal at low values
of the filling fraction.
Therefore, to the extent that the noncommutative dipole can be associated
to the Halperin--Kallin quasiexciton, the noncommutative formalism is
less and less relevant for practical calculations
 as we consider larger wavelengths for the density
fluctuations.

\newsec{Conclusions}

\noindent

We have analyzed the prospects for using techniques from NCFT, notably
perturbation theory with Moyal products, as a calculational technique for
the physics of density fluctuations of fractional Hall fluids.
The basic hypothesis of \refs\rsus\ is that promoting the electrons to
`D-brane' objects, i.e. enlarging the statistics group from $S_N$ to
$U(N)$, is a new and interesting way of performing the `flux-attachment'
transformation  that is at the basis of most effective field theory
treatments of the QHE \refs\rcsqh.  At the same time, the noncommutative fluid is
`granular' in character, with lenght scale $\sqrt{\theta}$,
 and we expect to see characteristic effects
of this granularity at wavelengths of $\CO(\sqrt{\theta})$.

We have seen that a straightforward application of the weak-field
expansion and noncommutative perturbation theory fails to capture
the physics due to strong UV/IR mixing, yielding infrared divergences
that destroy the incompressibility of the Hall fluid.

We consider unlikely
 the possibility that the UV/IR divergences should be interpreted
in terms of the Wigner-crystal phase transition. A standard
Wilsonian picture of the effective action suggests that the UV/IR divergences
are artifacts of the weak-field expansion and such simple effective actions
should be defined with an explicit ultraviolet cutoff at the scale
$\Lambda \sim 1/\sqrt{\theta}$. Under these conditions, the NCFT approach
to the  spectrum of density fluctuations at very long wavelengths
 is not qualitatively
different from that of ordinary hidrodynamics.

Therefore, in order to test the hypothesis of \refs\rsus\ at the level of the
bulk dynamics we must work at wavelengths of $\CO(\sqrt{\theta})$ or smaller,
and go beyond the weak-field expansion.
In this context, we have noticed that the geometrical structure
 of the Halperin--Kallin
quasiexciton exactly matches that of a
 noncommutative dipole of length $\theta |{\bf p}|$. We regard this
as really convincing evidence that the basic idea of \refs\rsus\ is
actually correct. Thus, it is tempting
to conjecture that the open Wilson lines are interpolating fields for the
quasiexcitons, a kind of non-local `vertex operators' for bound states
of quasiparticle solitons. In proving such a conjecture, it is likely that
the direct Coulomb interaction \coulp\
 must be kept exactly, without any further dynamical approximations.

\vskip1.0cm

\noindent {\bf Acknowledgements}

\vskip 0.1cm

\noindent

We are indebted to Luis Alvarez-Gaum\'e,
  C\'esar G\'omez,
Karl Landsteiner,  Esperanza L\'opez,  Alfonso Ramallo and
Miguel A. V\'azquez-Mozo 
 for  useful discussions.
A. P.  would also like to thank the CERN Theory Division for hospitality.
The  work of A. P. was
supported in part by DGICYT under grant PB96-0960,  by CICYT under
grant  AEN99-0589-CO2 and by Xunta de Galicia under  grant
PGIDT00-PXI-20609.

\appendix{A}{$\,$}

\noindent

In the evaluation of the one-loop
 contribution to the gauge boson self-energy
 in noncommutative Maxwell--Chern--Simons theory,
 one finds the following expression (notice
 that we are working on euclidean space):
\eqn\apauno{
\Pi_{\mu\nu}^{(1)} = {M\over k}\,
\int {d^3 q \over (2\pi)^3}
\, {P_{\mu\nu} (q,p,M) \over q^2 (q-p)^2
(q^2+M^2)((q-p)^2+M^2)}\,\,\sin^2 \left[ {\tp  q \over 2} \right]\,\,,}
where $P_{\mu\nu}$ is a complicated polynomial of momentum dimension six,
which comes from the sum of the three contributing one-loop diagrams,
 including the one with the ghost loop
 (for more details, see \refs\rpr\ and \refs\rshift.)
 Because of gauge invariance, \apauno\ must be of the form \tens.

The planar and nonplanar contributions can be separated performing the
usual substitution:
\eqn\apados{
\sin^2 \left[ {\tp q \over 2} \right] = {1 \over 2}\,(1 - \cos \tp  q)\,\,.}
Since the physical region is
 $|p| \ll M$, we are interested in a low momentum expansion in 
powers of $|p|/M$.
 Let us first concentrate on the infrared part of the integral.
 At first sight, it seems that it could be divergent as $|p| \rightarrow 0$.
 Nevertheless, it was rigorously proven by Pisarski and Rao that the
 infrared divergences cancel out. In fact, they exactly calculated the planar
 part of this integral, obtaining $\Pi_o^{(1)}$ and $\Pi_e^{(1)}$ as analytic
 functions of $|p|/ M$. This analyticity cannot be spoiled in the
 non-planar integral, as in this region the phase is slowly varying and it
 can be expanded in powers of $\tp  q$.
 With each power, the small $p$ contribution decreases.

For the purpose of
calculating the leading terms of the ultraviolet contribution
 to the nonplanar term of \apauno, we can set $|p| = 0$,
 while keeping $\tp$ on the phase factor,
which can lead to the UV/IR mixing, as it is well known.
Then, taking the explicit form of $P_{\mu\nu}$ in this limit, we get:
\eqn\apatres{
2\pi {M \over k}\,\int {d^3 q \over (2\pi)^3} \,
{2\,q_{\mu}q_{\nu}-\delta_{\mu\nu}\,M^2-\delta_{\mu\nu}\,q^2 \over
 (q^2+M^2)^2}\,e^{i\tp\,q}\,\,.}
Notice that the terms that
 would appear in \apatres\
 if we had not taken $p=0$ in the numerator are negligible at small $p$,
 as they would increase the powers of $p$ and decrease the powers of $q$
(thereby decreasing the powers of $1 / |\tp|$ in the final result.)

The integral \apatres\ can be easily computed taking derivatives with
 respect to the components of $\tp$ in the equality:
\eqn\apacuatro{
\int {d^3 q \over (2\pi)^3} \, {e^{i\tp\,q} \over (q^2+M^2)^2} =
{e^{-|\tp | M} \over 8 \pi M}.}

After a short calculation one obtains
\eqn\apacinco{
\Pi_{\mu\nu}^{(1)}(p \rightarrow 0)_{\rm nonplanar} =
-{M \over 2k}\,{\tp_\mu\tp_\nu \over  |\tp|^3}(1+|\tp|M)e^{-|\tp | M} =
-{M \over 2k}\,{\tp_\mu\tp_\nu \over  |\tp|^3}
\left(1-{(|\tp | M)^2 \over 2}+\ldots\right)}

Despite the existence of the Chern--Simons mass,
the UV/IR effect renders the dispersion relation singular in the infrared.
At small $p$, we get the following dependences
 of the one-loop correction to the quantities appearing
in the dispersion relation \diss:
\eqn\apaseis{
\Pi_o^{(1)}\propto {1 \over k}\, |p|^0\,, \quad
\Pi_e^{(1)}\propto {1 \over k}\, |p|^0\,, \quad
|\tp |^2\,\Pi_{nc}^{(1)}\propto {M \over 2k}\,{1\over |\tp |}\,\,,}
where we have returned to the Lorentzian signature $(+--)$ used in the
main text.

\appendix{B}{$\,$}

\noindent

It is known that a diagram with a fermion loop (with the fermion
 transforming in the adjoint representation) leads to an
 UV/IR divergence of the same type as a vector,
 but with opposite sign in the photon self-energy.
 The coefficient of the divergent term is independent of the fermion mass
 $m_f$, which works like a gauge-invariant   Pauli--Villars regulator.
 Thus, if a Majorana fermion is included in the theory so
 the number of fermionic degrees of freedom equals the number
 of bosonic ones, the cutoff will eliminate the divergence in the dispersion
 relation \diss.

Our goal in this appendix is to show explicitly that the gap in the dispersion
 relation at $|{\bf p}|=0$
 is recovered (does not depend substantially on $m_f$),
as was stated by general arguments in the main text.
From the fermion diagram, one gets a one-loop contribution:
\eqn\apbuno{
\Pi_{\mu\nu}^f\,\sim\,{M \over k}\,\int {d^3 q \over (2\pi)^3}\,{\rm tr}
\left[ \gamma^\mu \,{q\!\!\!\slash-p\!\!\!\slash-m_f \over (q-p)^2+m_f^2}\,
\gamma^\nu \,{q\!\!\!\slash-m_f \over q^2+m_f^2} \right]\,
\sin^2 \left[ {\tp q \over 2} \right]\,\,.}

It was proved in \refs\rchucs\ that the planar part of the integral just induces a
 Chern--Simons term, renormalizing the original Chern--Simons coefficient
$k\rightarrow k \pm {1 \over 2}$. Thus, let us concentrate in the nonplanar part.
 All the integrals needed to compute it can be obtained by differentiation from:
\eqn\apbdos{
{\rm I}
 = \int {d^3 q \over (2\pi)^3}\,
 {e^{i\tp\,q} \over ((q-p)^2+m_f^2)(q^2+m_f^2)}\,\,.}
Introducing Feynman parameters and defining:
\eqn\apbtres{
\Omega = -x(x-1)p^2+m_f^2\,\,,}
we have:
\eqn\apbcuatro{\eqalign{
{\rm I}&=\int_0^1\,dx\,e^{i\tp\,p\,x}\,\int {d^3 q \over (2\pi)^3}\,
{e^{i\tp\,q} \over (q^2+\Omega)^2}\,=
\int_0^1\,dx\,e^{i\tp\,p\,x}\,
\int_0^{\infty}d\alpha \,\alpha e^{-\alpha \Omega}
\int {d^3 q \over (2\pi)^3}\,e^{i\tp\,q-\alpha q^2} \cr
&={1 \over 8 \pi^{3\over 2}}\int_0^1\,dx\,e^{i\tp\,p\,x}\,
\int_0^{\infty}d\alpha \,
 {1 \over \sqrt{\alpha}}\,e^{-\alpha \Omega-{|\tp|^2 \over 4 \alpha}}=
{1 \over 8\pi}\,\int_0^1\,dx\,e^{i\tp\,p\,x}\,
{e^{-|\tp| \sqrt{\Omega}} \over\sqrt{\Omega}}\,\,.}}
Notice that the condition $\tp\cdot p=0$
cannot be imposed until no more derivatives are to be done,
 since we must consider $p$, $\tp$ to be independent parameters.

In order to obtain the zero-momentum contribution to the mass gap,
$$
{e^{-|\tp| \sqrt{\Omega}} \over\sqrt{\Omega}}
$$
 can be expanded in powers of $p^2 / m_f^2$:
\eqn\apbcinco{\eqalign{
{\rm I}=&{e^{-|\tp|m_f }\over 16 \pi m_f (\tp p)^3}\,{\Big [}-2i(-1+e^{i\tp p})(\tp p)^2
 \cr
&+
(1+|\tp|m_f)\left(-2i+(\tp p)+e^{i\tp p}(2i+\tp p)\right){p^2 \over m_f^2}+
\CO(
p^4 / m_f^4) {\Big ]}.}}
Differentiating this expression and inserting the results in \apbuno,
 one gets, after some calculation (and  rotating back to Lorentzian
signature):
$$
\Pi_o^f(p\rightarrow 0)\propto {M \over k}\,
{1 \over m_f}\,e^{-|\tp|m_f} \,\,,
$$
$$
\Pi_e^f(p\rightarrow 0)\propto {M \over k}\,
{1 \over m_f}\,e^{-|\tp|m_f} \,\,,
$$
\eqn\apbocho{
|\tp |^2\,\Pi_{nc}^f(p\rightarrow 0)=
 -{M \over k}\,{1 \over 2 |\tp |}\,e^{-|\tp|m_f}(1+|\tp|m_f)
= -{M \over k}\,{1 \over 2 |\tp |}+\CO(|\tp|)
\,\,,}
while the gauge non-invariant terms naturally vanish.
So \apbocho\ cancels out the divergence \apaseis, while the vanishing of
the coefficient of $|\tp|^0$ in the Taylor expansion of
\apbocho\ makes stable the value of the gap at
  $|p|=0$ when increasing the mass of the fermion.

As the mass of the fermion goes to infinity, the dispersion relation
 with cutoff must approach the one without it,
 so we must have curves that are qualitatively like the ones showed 
in Fig. 2.

\listrefs

\bye